\documentclass{jot}
\usepackage[nohyperlinks,printonlyused,nolist]{acronym}
\usepackage[most]{tcolorbox}
\usepackage{xcolor}
\usepackage{float} 
\usepackage{capt-of}  
\usepackage[acronym]{glossaries}
\usepackage{minted}
\usepackage{booktabs}
\usepackage{siunitx}
\usepackage{tabularx}
\usepackage{indentfirst} 
\usepackage{todonotes}
\usepackage{booktabs}     
\usepackage{threeparttable} 
\usepackage{ragged2e}
\usepackage{epstopdf}
\usepackage[table]{xcolor}

\usepackage{eso-pic}
\AtBeginDocument{%
  \AddToShipoutPictureBG*{%
    \AtPageUpperLeft{%
      \raisebox{-3.5cm}{
        \hspace{1.5cm}
        \parbox{0.85\textwidth}{%
          \small\itshape\color{red} This is a Preprint. The final version is published in the Journal of Object Technology. 
        }%
      }%
    }%
  }%
}

\definecolor{siggray}{gray}{0.85}   
\newcommand{\sig}[1]{\cellcolor{siggray}#1}

\definecolor{myblue}{RGB}{0, 0, 255}
\definecolor{myred}{RGB}{255, 0, 0}
\definecolor{mygreen}{RGB}{0, 128, 0}
\definecolor{eclipsepurple}{RGB}{127, 0, 85}

\jotdetails{ 
volume=25, 
number=3, 
articleno=a42, 
year=2026, 
license=ccby 
}

\newtcolorbox{promptbox}[2][]{%
  enhanced,
  floatplacement=H,          
  colback=blue!3,
  colframe=blue!50!black,
  arc=1mm, boxrule=0.4pt,
  left=10pt, right=0pt, top=0pt, bottom=0pt,
  title={#2}, fonttitle=\bfseries\scriptsize,
  before=\par\noindent\begin{minipage}{\columnwidth},
  after=\end{minipage}\par,
  #1
}

\lstdefinestyle{reactionsstyletiny}{
  basicstyle=\scriptsize\ttfamily,
  commentstyle=\color{mygreen},
  keywordstyle=\color{myblue},
  stringstyle=\color{myred},
  breaklines=true,
  morekeywords={reactions, changes,reaction, routine, execute, action, after, call, require, absence, action, and, initialize, correspondence, between, tagged, with, match, inserted,  retrieve, actions, corresponding, update, check, create}, 
  morekeywords=[2]{import, element, created,  in, root, new, as, in, of,val, to},
  keywordstyle=[2]\color{eclipsepurple},
  moredelim=**[is][\color{myblue}]{@1}{@},
  moredelim=**[is][\color{myred}]{@2}{@},
}

\tcbset{
  finding/.style={
    breakable,
    enhanced,
    colback=black!4,      
    colframe=black!20,    
    boxrule=0.3pt,        
    arc=2pt,             
    left=4pt,right=4pt,top=3pt,bottom=3pt, 
    before skip=6pt, after skip=6pt,      
    width=\columnwidth,  
  },
  findingtitle/.style={
    fonttitle=\bfseries,
    coltitle=black,
    title={#1},
    attach boxed title to top left={yshift*=-1pt,xshift=2pt},
    boxed title style={colback=black!6, frame empty, interior style={opacity=0.9}},
  }
}

\newtcolorbox{findingbox}[2][]{finding, findingtitle={#2}, #1}

\newtcolorbox{findingnote}[1][]{finding, #1}

\articletype{regular}

\title{LLM4MTLs: Automated Generation and Empirical Evaluation of Model Transformation Languages}

\def\anonymous{0}

\newcommand{\authorKit}[2]{
\if\anonymous1
  \author[#2]{Anonymous Author}
  \affil[#2]{Anonymous University}
\else
  \author[#2]{#1}

\fi
}

\authorKit{Bowen Jiang}{$\ast$}
\authorKit{Nathan Hagel}{$\ast$}
\author[$\dagger$]{Haowei Cheng}
\authorKit{Benedikt Jutz}{$\ast$}
\authorKit{Arne Lange}{$\ast$}
\authorKit{Weixing Zhang}{$\ast$}
\authorKit{Rahul Sharma}{$\ast$}
\authorKit{Ralf Reussner}{$\ast$}
\authorKit{Anne Koziolek}{$\ast$}

\affil[$\ast$]{Karlsruhe Institute of Technology, Germany}
\affil[$\dagger$]{Waseda University, Japan}

\runningtitle{LLM4MTLs: Automated Generation and Empirical Evaluation of Model Transformation Languages}
\runningauthor{Jiang \textit{et al.}}

\begin{abstract}
Model transformation languages (MTLs) are domain-specific languages used to transform models conforming to a given metamodel into other models, including textual models such as source code. Developing correct model transformations in these languages is challenging and requires both language-specific and domain knowledge, creating a need for automated assistance and thus motivating the use of large language models (LLMs) for MTL code generation. However, due to the limited availability of training data and executable examples, LLM-generated MTL code is often not syntactically valid or semantically usable out of the box. 
This paper presents \textit{LLM4MTLs}, an automated workflow for constructing and comparing prompting strategies for
LLM-generated MTL code,
together with an evaluation suite and an empirical evaluation.
The workflow systematically explores prompt constructions combining few-shot prompting, grammar prompting, and helper methods inclusion, and evaluates them using both syntactic and semantic metrics. We construct an evaluation suite spanning four MTLs (ATL, ETL, QVTo, and the Reactions language) with executable reference scripts and manually written test suites, and evaluate across three LLMs. 
We find that few-shot prompting consistently improves syntactic quality across all four MTLs while gains in semantic correctness are uneven and language-dependent. For ATL, Pass@1 remains unchanged across all strategies and models, indicating that few-shot prompting improves surface-level syntax more readily than deep transformation semantics. 
Grammar prompting stabilizes code generation when combined with few-shot examples, but in isolation, it can be ineffective or even counterproductive for certain model–language combinations. 
Furthermore, including helper methods in the prompt as a complementary amplifier can be beneficial. Finally, LLM Model choice influences syntactic correctness and similarity for certain MTLs, particularly ETL and QVTo, while its influence on semantic correctness remains limited across all MTLs.

\end{abstract}

\begin{document}

\maketitle
\urlstyle{rm}


\begin{acronym}
    \acro{LLM}{Large Language Model}
    \acro{DSL}{Domain Specific Language}
    \acro{MTL}{Model Transformation Language}
    \acro{ATL}{Atlas Transformation Language}
    \acro{MDSD}{Model-Driven Software Development}
    \acro{QVT}{Query/View/Transformation}
    \acro{MBE}{Model-Based Engineering}
    \acro{MOF}{Meta Object Facility}
    \acro{OOP}{Object-Oriented Programming}
    \acro{OCL}{Object Constraint Language}
    \acro{GPL}{General Purpose Language}
    \acro{M2M}{Model-to-Model}
    \acro{M2T}{Model-to-Text}
    \acro{VPDL}{ViewPoint Definition Language}
    \acro{UML}{Unified Modeling Language}
    \acro{AAS}{Asset Administration Shell}
    \acro{MTBE}{Model Transformation by Example}
    \acro{T2M}{Text-to-Model}
    \acro{VSUM}{Virtual Single Underlying Model}
    \acro{ChrF}{Character n-gram F-score}
    \acro{QVTo}{QVT-Operational Language}
    \acro{ETL}{Epsilon Tranmsformation Language}
    \acro{LoC}{Lines of Code}
    \acro{EBNF}{extended Backus–Naur form}
    \acro{RAG}{Retrieval-augmented generation}
\end{acronym}

\setlength{\parindent}{2em}
\setlength{\parskip}{0pt}

\section{Introduction}
\label{sec:introduction}
\acp{LLM} have now been firmly established in generating source code from natural language descriptions~\cite{joelSurveyLLMbasedCode2025, DBLP:journals/corr/abs-2510-26275}.
While code generation is mostly done for general-purpose languages such as Python or Java, \acp{LLM} are also increasingly being explored in specialized domains, including \ac{MDSD}. For example, \acp{LLM} are used to generate model instances \cite{garaccioneComparisonDifferentLarge2025} and to verify the consistency of models~\cite{wimmerLLMSForMerging}.

\emph{Model Transformations} are central to \ac{MDSD}~\cite{mensTaxonomyModelTransformation2006}. In a model transformation, a set of source models is transformed into a set of target models for purposes of analysis, simplification, or generation of runnable \ac{GPL} code, like Java or C++.
Transformations consist of a set of transformation rules written in an \ac{MTL} which provides optimized syntax and language features for definition. Examples of \ac{MTL}  include the \ac{ATL}~\cite{jouaultATLModelTransformation2008},  \ac{ETL}~\cite{kolovosEpsilonPlayground2022}, and Acceleo~\cite{AcceleoHome}. 
Nevertheless, creating transformations between models can be a timely and challenging task, requiring knowledge not only about the \ac{MTL}, but also about both source and target models, or even about the domains where those models originate \cite{anastasakis2007uml2alloy}. 

Given the challenge of manually developing correct transformations, 
it is natural to investigate 
whether LLMs can assist in generating MTL code.
However, applying \acp{LLM} to this task 
presents unique challenges.
Firstly, \acp{MTL} are \acp{DSL} that have limited representation in the training data of \acp{LLM}, which implies that \acp{LLM} typically lacks knowledge of MTL-specific syntax and idioms~\cite{joel2024survey}.
Secondly, \acp{MTL} and model transformations depend on metamodel semantics concerning the source and the target metamodel, which are the rules and definitions of how concrete model instances can be defined. Although metamodels offer the necessary context, they also introduce specialized semantic constraints that require domain-specific reasoning, which remains challenging for \ac{LLM} ~\cite{joel2024survey}.

Out-of-the-box prompting frequently leads to the generation of transformation code with syntactic or semantic errors, even when metamodel information is provided. The state of the art shows LLM-generated transformations succeed only in simple scenarios, but fail in more complex tasks~\cite{buchmann_prompting_2024}. In practice, finding effective prompts becomes a manual trial-and-error process~\cite{ye2023prompt_trial_and_error, pryzant2023automatic_trial_and_error}. This leaves two research gaps. Firstly, there is no systematic and reproducible approach for optimizing and evaluating LLM-based \ac{MTL} code generation. Secondly, there is a lack of publicly available evaluation suites with reference transformations and executable test suites across multiple \acp{MTL}~\cite{2026llm4mde, burguenoAutomationModelDrivenEngineering2025}.
This paper addresses both gaps by investigating the following \textbf{Research Question}:
\textbf{How can we systematically construct and evaluate prompting strategies for \ac{LLM}-based \ac{MTL} code generation, and to what extent do such strategies improve syntactic and semantic quality across different \acp{MTL}?}




We introduce \textit{LLM4MTLs}, an automated, metric-driven workflow that standardizes prompt construction, code generation, and evaluation for \ac{MTL} code generation. 
The workflow targets measurable syntactic and semantic outcomes and provides a reproducible means of exploring and comparing different prompt configurations.
We further construct an evaluation suite of transformation examples spanning four \acp{MTL}: \ac{ATL}~\cite{atl2018syntax}, \ac{ETL}~\cite{kolovosEpsilonPlayground2022}, \ac{QVTo}~\cite{MOFQueryViewc}, and the Reactions language of the Vitruvius framework~\cite{klareEnablingConsistencyViewbased2021}. The evaluation suite includes reference transformation scripts, metamodels, and manually written test suites for both syntactic and semantic validation.
Using this workflow and evaluation suite, we conduct an empirical evaluation of generated \ac{MTL} code combining prompting strategies, including few-shot prompting, grammar prompting, and Xtext-based language-specific helper method inclusion across three \acp{LLM}: \citet{gpt5_model_nodate}, \citet{noauthor_gemini_nodate}, and~\citet{noauthor_claude_nodate}. 
The workflow is realized using n8n workflow automation~\cite{N8nioAIWorkflow}; all components are containerized, and all artifacts are provided in a replication package~\cite{jiangLLMbasedCodeGeneration2025}.
In summary, this paper makes the following three contributions:
\begin{itemize}[noitemsep, topsep=2pt]
  \item An evaluation suite of 47 transformation examples across four \acp{MTL}, consisting of executable reference scripts, metamodels, and test suites for automated validation.
  \item An automated workflow for prompt generation, \ac{MTL} code generation, and metric-driven evaluation.
  \item An empirical evaluation of our workflow across four \acp{MTL} and three \acp{LLM}, offering systematic evidence on how different prompting strategies and \acp{LLM} choices affect the syntactic quality and semantic correctness of generated \acp{MTL} code.
\end{itemize}

The rest of this paper is structured as follows.
In \autoref{sec:background}, we provide background on \acp{MTL} as our application domain.
In \autoref{sec:relatedWork}, we survey related work with regards to using \acp{LLM} for generating code for \acp{MTL}, and other low-resource languages, such as other \acp{DSL}.
In \autoref{sec:approach}, we provide an overview of our optimization workflow and describe the implementation. We present our evaluation suite and the results of our evaluation afterwards in \autoref{sec:evaluationandresults}.
We discuss the findings, limitations, and threats to validity in \autoref{sec:discussion}, and
finally, we conclude the paper in \autoref{sec:conclusion}.
\section{Background}
\label{sec:background}

\subsection{\acl{MDSD}}
\ac{MDSD} is the definition and use of software models, such as \ac{UML} class or sequence diagrams, to create runnable software systems. It advances the concept of \ac{MBE} further.
There, the same models are used for design and communication, but developers do not generate runnable program code from these models~\cite{brambillaModelDrivenSoftwareEngineering2017}.

In the context of \ac{MDSD}, the term "model" usually does not refer to machine learning models, such as \acp{LLM}.
Instead, models conform to structures similar to those in \ac{OOP}.
To be more exact, a model must conform to a \emph{metamodel}, which is another model that defines a set of valid models or \emph{instances}~\cite{brambillaModelDrivenSoftwareEngineering2017}.
Metamodels consist of an abstract syntax that describes the structure of models, at least one concrete syntax to express models with (such as grammars for textual models), and the semantics for models~\cite{brambillaModelDrivenSoftwareEngineering2017}.
Commonly used standards for metamodelling include the \ac{MOF} to define the abstract syntax~\cite{MetaObjectFacility2019}, and the \ac{OCL} to describe additional restrictions on models through their semantics~\cite{ObjectConstraintLanguage2014}.
One of the main purposes of \ac{MDSD} is to automate software development.
This can be done either by interpreting the models directly, or to generate code from them by applying model transformations on them \cite[Sect. 3.1]{brambillaModelDrivenSoftwareEngineering2017}.

\subsection{Model Transformations}
Model transformations automatically convert a set of \emph{source models} into a set of \emph{target models}.
They require a \emph{definition} that describes how this conversion occurs.
Such a definition is a set of \emph{transformation rules}. 
Each rule describes how to translate constructs of the source- into constructs of the target metamodel~\cite{DBLP:books/daglib/0010435}. 
Since \ac{MDSD} treats code as models, model transformations can also take code as source and target models.
In fact, generating running code from models is a main purpose of model transformations~\cite{mensTaxonomyModelTransformation2006}.
Following~\citeauthor{brambillaModelDrivenSoftwareEngineering2017}, we call model transformations with code as source model \ac{T2M} transformations, transformations with code as target model \ac{M2T} transformations, and transformations only with non-code models \ac{M2M} transformations.

Model transformations are usually written with their own \acp{DSL}, such as the \acl{ATL}~\cite{jouaultATLModelTransformation2008}, or the \ac{QVT} family~\cite{MOFQueryViewc} of transformation languages.
Of course, transformations can also be written in a \ac{GPL} such as Java.
However, users of \acp{MTL} find that these specialized languages offer numerous features that increase their comprehensibility and productivity, among other quality attributes~\cite{hoppnerAdvantagesDisadvantagesDedicated2022}.

\subsection{View-based Modeling -- The \acs{VSUM} approach}
\label{ssec:applicationDomain}

View-Based Modeling~\cite{atkinson_orthographic_2010} is an approach in which views are tailored to the needs of individual stakeholders, thereby reducing the perceived complexity of the overall system under investigation.
At its core lie \acp{VSUM}~\cite{klareEnablingConsistencyViewbased2021}, which consolidate all essential model information and expose views that contain only relevant aspects while omitting extraneous details.
The Vitruvius framework implements the \ac{VSUM} approach and provides the \emph{Reactions Language}~\cite{klareBuildingTransformationNetworks2021}, a domain-specific language for specifying consistency preservation rules.
Each rule reacts to a designated change in one view and propagates the resulting modifications throughout the system, ensuring that all constituent models remain mutually consistent at all times.
\section{Related Work}
\label{sec:relatedWork}


\begin{figure*}[tb]
\includegraphics[width=\linewidth]{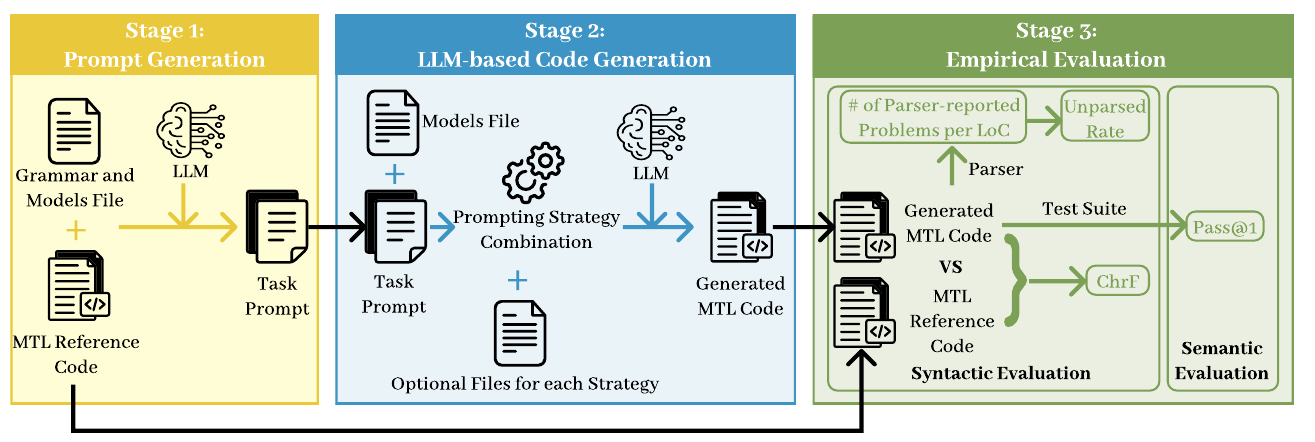}
\caption{Overview of the LLM4MTLs Approach.}
\label{fig:general_approach}
\end{figure*}

\begin{figure}[tb]
  \centering
  \includegraphics[width=0.9\columnwidth]{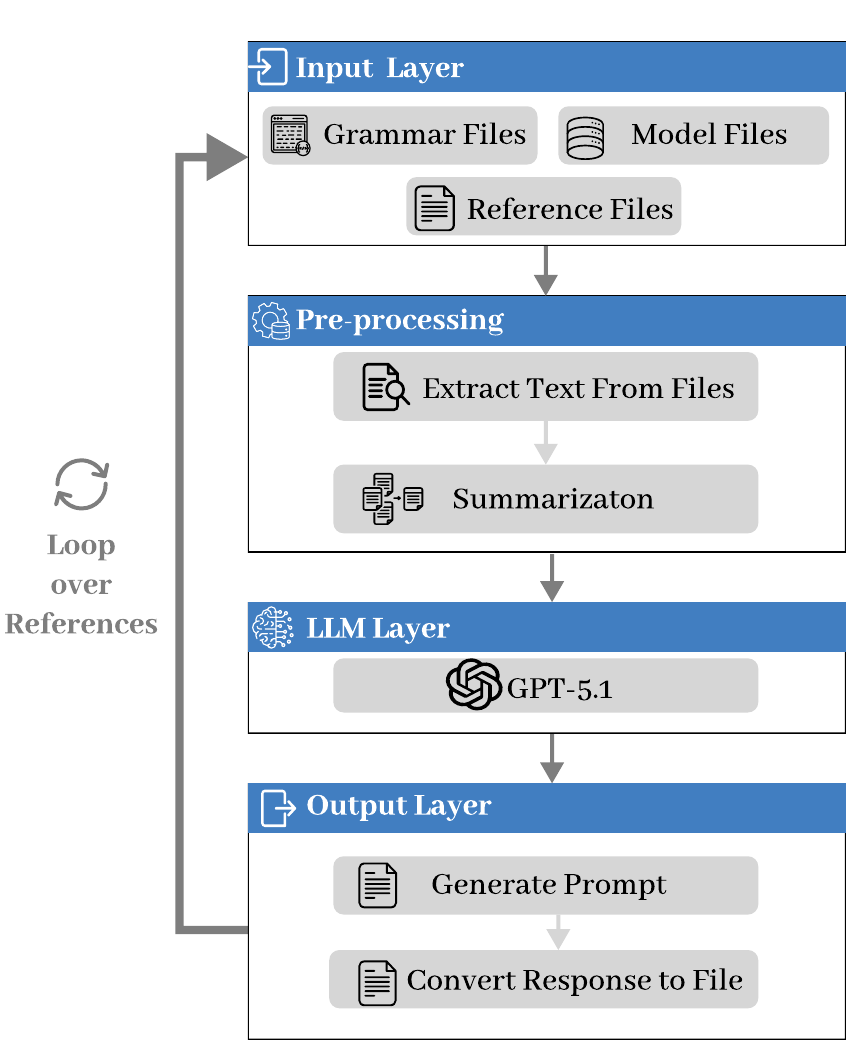}
  \caption{Detailed workflow of Prompt Generation.}
  \label{fig:prompt_generation}
\end{figure}


\subsection{Applying \acp{LLM} for model transformations}
\label{ssec:llmsForModelTransformation}
\acp{LLM} is finding widespread application for model transformation tasks.
They are used both to generate model transformation rules and to do the actual transformation themselves~\cite{zhang2026llm4mde}.
Therefore, we present works from both research areas here.

\citeauthor{buchmann_prompting_2024} prompted \ac{M2M} transformations in Java with ChatGPT for the Families-to-Persons example \cite{anjorinFamiliesPersonsCase}, a commonly used case study in the field of model transformations \cite{buchmann_prompting_2024}.
The generated model transformations worked well in the batch case, i.e., converting one source model to a target model.
However, prompting for incremental transformations, which only transform the changed part of a model, failed with either compilation errors or failing tests. 
\citeauthor{pontesmirandaInContextLLMBasedApproach2024a} evaluated the prompting of view definitions in the \ac{VPDL} of the EMF Views framework~\cite{pontesmirandaInContextLLMBasedApproach2024a}.
View definitions transform interrelated models with different metamodels into another model, or view, that combines the information of the underlying models.
The authors include view descriptions and metamodels converted to PlantUML for the prompt, and use chain-of-thought prompting.
They ensure the syntactical correctness of the output by parsing and reprompting the \ac{LLM}.
The performance of \acp{LLM} is limited at best, however, since the generated view definitions capture only a subset of the relations between models and their views (Precision $\in [0, 0.58]$).

In~\cite{cibrian_automating_2025}, the authors proposed an agent-based \ac{LLM} pipeline that transforms Modelica models to the SysML v2 representation. Their approach delivered promising results with a mean precision of 89.05\%. The usage of \acp{LLM} also caused some problems; it introduced inconsistencies in the required validation loops to correct the mistakes made by the \ac{LLM}. Their evaluation also lacked the semantic verification and was based only on structure. 


\subsection{\acp{LLM} in other \ac{MDSD} areas}
\label{ssec:llmsForGeneralMDSD}
In~\cite{camara_assessment_2023}, the authors investigated the generation of \ac{UML} diagrams via prompts to ChatGPT. The diagrams created or assisted by this method were generally correct, but contained syntactical errors. Some modeling concepts that \ac{UML} supports, such as multiple inheritance or integrity constraints, are not covered by ChatGPT. PlantUML seems to work best notation-wise than other modeling notations/languages to create \ac{UML} diagrams, e.g., USE (the UML-based Specification Environment). Rather than avoiding the \ac{LLM} assisted modeling, the authors recommend improving the quality and quantity of publicly available modeling artifacts so that \acp{LLM} have more data to train on.

Generating \ac{OCL} constraints with the help of \acp{LLM} is the focus of~\cite{abukhalafPathOCLPathBasedPrompt2024a} and works reasonably well. PathOCL, the name of the presented method, combines prompting techniques with simple path coverage that subsets the \ac{UML} class diagram to only relevant portions of the model. The results are promising and need further investigation and optimization to be used reliably. This work also saw better and more publicly available \ac{MDSD} artifacts as the limiting factor to better performance of \acp{LLM} in this task.
Beyond model and constraint generation, LLMs have also been applied to instance generation in the context of language evolution.
Zhang et al.~\cite{zhang2025leveraging, zhang2026leveraging} explored using LLMs to support grammar-instance co-evolution for Xtext-based textual DSLs, exploring the potential of LLMs to preserve auxiliary information such as comments and formatting. 
Hagel et al.~\cite{hagel2024turning} analyzed how \acp{LLM} can be applied to generate models using a textual DSL in a model-based low-code tool. 
Results show that the \ac{LLM} was capable of generating a proprietary, to the \ac{LLM} unknown DSL, and a user study showed that task completion time could be reduced.
While those studies apply LLMs to DSL-related code generation, their work focuses on instance evolution, whereas ours addresses MTL generation from natural language.
In Eisenberg et al.~\cite{wimmerLLMSForMerging} the authors apply \acp{LLM} to detect conflicts in versioned models and automatically resolve them using \acp{LLM}. 
In \cite{hagel2025towards} the authors discuss the usage of \acp{LLM} for explicit model consistency based on natural language or formal consistency rules.
Our study addresses MTL generation, including the generation of the Reactions language for automated consistency maintenance, rather than applying LLMs to resolve inconsistencies directly.
Besides, \acp{LLM} have also been explored for adjacent domain-specific and formal modelling tasks, including self-adaptive systems~\cite{DBLP:journals/taas/LiZLWJT24,DBLP:conf/seams/LiZLWJT24} and discrete controller synthesis~\cite{DBLP:journals/corr/abs-2512-07261}.

\section{LLM4MTLs - An Automated Workflow}
\label{sec:approach}

We present \textit{LLM4MTLs}, an automated workflow for improving the reliability of \ac{LLM}-generated \ac{MTL} code through systematic prompt composition and metric-driven evaluation.
The workflow targets measurable syntactic and semantic outcomes and provides a reproducible means of exploring and comparing different prompt configurations.
An overview is depicted in \autoref{fig:general_approach}, where the three main stages are distinguished by color.
In the \textbf{Prompt Generation} stage (yellow), \textit{task prompts} are automatically derived from existing \ac{MTL} reference implementations to address the lack of structured evaluation datasets (\autoref{sc:prompt_generation}).
In the \textbf{Code Generation} stage (blue), a range of prompting strategies and their combinations are applied to generate \ac{MTL} code from these task prompts (\autoref{sc:llm_code_generation}).
In the \textbf{Empirical Evaluation} stage (green), the generated code is assessed against the reference implementations using syntactic similarity, syntactic correctness, and semantic correctness metrics (\autoref{sec:evaluationandresults}).
Here, \textit{Grammar Files} refer to \ac{EBNF} grammars of the target \ac{MTL}; \textit{Model Files} refer to \textbf{metamodel artifacts} for both Prompt Generation and Code Generation. Test suite additionally loads concrete \textbf{model instances}.

Our workflow emphasizes automation and thus supports efficient repetition and fast feedback cycles when searching for an optimal prompting strategy configuration.
Strategies and the associated resources can be adjusted to accommodate different target \acp{MTL}.
 To ensure reproducibility, all required components---including the workflow definition, model files~(metamodel artifacts), \ac{MTL} grammars, and code snippets---are bundled within containers.
Our workflow is implemented using n8n~\cite{N8nioAIWorkflow}, a self-hosted low-code tool that enables \ac{LLM} workflows to be visually constructed and edited.
To support traceability and systematic comparison across prompt configurations, all prompts and \ac{LLM}-generated artifacts are persisted.
The complete implementation, including workflow definitions and evaluation artifacts, is made available as a replication package~\cite{jiangLLMbasedCodeGeneration2025}.


\subsection{Prompt Generation}
\label{sc:prompt_generation}
We enable an automated dataset construction across \acp{MTL} by employing an \ac{LLM} to reverse-engineer \textit{task prompts} from existing reference code snippets, formulated in a developer-to-AI voice.
The snippets provided enable the \ac{LLM} to capture context and intent even when it is unfamiliar with a given \ac{MTL} and is, consequently, incapable of generating high-quality code directly.
Minor imperfections in the generated \textit{task prompts} are acceptable, because human-authored prompts are rarely perfect too.
Crucially, by encoding additional contextual details directly in the prompt, our automated generation approach facilitates adaptation to further languages and broader example sets.
The generated \textit{task prompt}, together with the associated model files~(metamodel artifacts), serves, in the end, as input for baseline code generation.

\autoref{fig:prompt_generation} illustrates the prompt generation workflow.
Upon execution, the workflow reads the reference code snippets from a specified directory and iterates over each file.
To enrich the contextual information available to the \ac{LLM}, it optionally loads the corresponding model files~(metamodel artifacts) and grammar definitions.
The name of each reference file is preserved to maintain the association between the generated prompt and its source.
To ensure stylistic consistency, every query is prefixed with the system prompt shown in Listing \autoref{lst:system_prompt_gen_rl}, which instructs the \ac{LLM} to produce concise prompts that capture only the transformation intent and the relevant method names.
Although this system prompt is formulated for the Reactions language, it can be straightforwardly adapted to other \acp{MTL}.

\begin{promptbox}{System Prompt}
\begin{lstlisting}[breakindent=0pt]
You are acting as a senior developer who knows the Model Transformation Language: **Reactions Language** inside out. Your task is to reconstruct the NATURAL-LANGUAGE request that could have produced the code shown in the next message, as if a developer were asking an AI assistant to generate that code.
Rules
1. You MUST explicitly state the reaction and routine names.  
2. You MUST describe the intent on a high abstraction level.
3. DO NOT say 'model::component', 'component' is enough for the context.
   DO NOT say 'in the update block...', rather describe generally what happens in the routine and reaction.
   Do NOT mention 'snippet', 'reverse-engineer', or any analysis meta language.
4. When there are parts not included in the grammar, provide the method names.
5. Write in the *developer-to-AI* voice, e.g.  'Create a transformation [...] that...'.
6. Keep the request under **100 words**.
\end{lstlisting}
\end{promptbox}
\captionof{listing}{System Prompt for Prompt Generation (Reactions)}\label{lst:system_prompt_gen_rl}
\vspace{2pt}

The workflow pairs the system prompt with a user prompt that supplies the additional context required to generate correct transformation definitions.
Specifically, the user prompt contains the reference code, along with the extracted and summarized content of the associated model and grammar files.
A representative example---the \textsc{PersonToFamily} transformation for the Reactions language---is presented in Listings~\ref{lst:example_reaction_pg} and~\ref{lst:gen_prompt_rl}, illustrating how the intent of the original code is clearly conveyed in the generated prompt. 
The prompt generation workflow thereby expands the dataset with consistent, semantically aligned prompt--code pairs.
To ensure prompt quality, all generated prompts were manually reviewed for semantic correctness and fidelity to the intent of the original code.
This verification step is critical, as all subsequent strategy evaluations are predicated on these prompts.

\begin{promptbox}{Reactions Language}
\begin{lstlisting}[style=reactionsstyletiny]
[...]
reactions: familiesToPersons
in reaction to changes in families
execute actions in persons
reaction DeletedMember {
  after element families::Member deleted
  call deletePerson(affectedEObject)
}
routine deletePerson(families::Member member) {
  match {
    val person = retrieve persons::Person corresponding to member
    val family = retrieve families::Family corresponding to person
  }
  update {
    removeObject(person)
    removeCorrespondenceBetween(member, person)
    removeCorrespondenceBetween(family, person) 
  }
}
\end{lstlisting}
\end{promptbox}
\captionof{listing}{PersonToFamily.reactions Example}
\label{lst:example_reaction_pg}

\begin{promptbox}{Prompt}
\begin{lstlisting}[breakindent=0pt]
Create a transformation in the **Reactions Language** that synchronizes the *families* and *persons* models. 
Define a reaction named **DeletedMember** that triggers whenever a `Member` is deleted in the families model. This reaction should call a routine named **deletePerson**, which retrieves the corresponding `Person` and its `Family`, then removes the `Person` object and deletes all correspondences between the deleted `Member`, the `Person`, and the `Family`.
\end{lstlisting}
\end{promptbox}
\captionof{listing}{Generated Task Prompt}\label{lst:gen_prompt_rl}
\vspace{2pt}

\subsection{Code Generation}
\label{sc:llm_code_generation}

\begin{figure*}[]
  \centering  \includegraphics[width=0.85\textwidth]{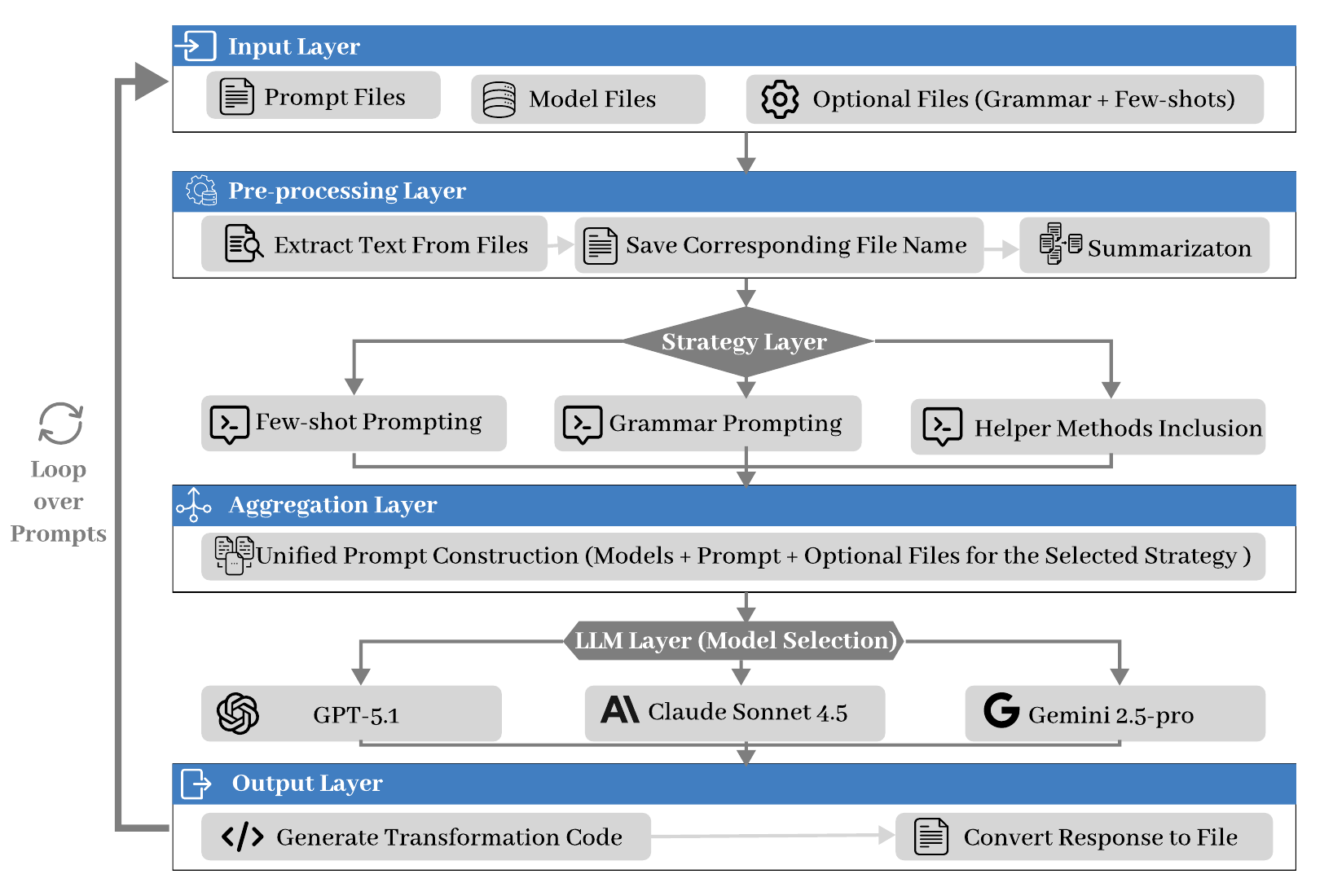}
  \caption{Detailed workflow of LLM-based MTL Code Generation}
  \label{fig:LLM-based-code-generation}
\end{figure*}
Compared to the prompt generation stage, the code generation workflow illustrated in \autoref{fig:LLM-based-code-generation} incorporates additional steps for optimization and execution.
The workflow first reads the task prompt files produced by the prompt generation stage, and extracts the task prompt text, the associated model files~(metamodel artifacts), and the name of the corresponding reference code snippet.
It then applies the prompting strategies selected by the user.

Three strategies are currently supported: (i)~\emph{few-shot prompting} (k=3), which augments the prompt with three representative reference examples of the target \ac{MTL}; (ii)~\emph{grammar prompting}, which appends an \ac{EBNF} grammar excerpt of the target \ac{MTL} to impose explicit syntactic constraints and guide the \ac{LLM} toward syntactically valid outputs~\cite{wang2023grammar}; and (iii)~\emph{helper method inclusion}, which supplies predefined utility functions to the \ac{LLM}'s context.
The third strategy can be used for \acp{MTL} that express key transformation constructs outside of their syntax.
This is the case for the Reactions language, where managing consistency happens within Java-defined methods, that need to be called within reactions.
The n8n workflow is designed to accommodate additional strategies through the straightforward inclusion of further files or resources in the prompt.

\begin{promptbox}{System Prompt}
\begin{lstlisting}[breakindent=0pt]
You are an expert developer for the **Reactions language** (model transformation DSL).  Your job is to translate the user's natural-language specification into a complete, syntactically valid .reactions file.
Rules  
1. Follow the DSL grammar exactly (imports, transformation block, reactions, routines, guards, create/update sections, persistence paths, correspondence links, etc.).  
2. Use the transformation / reaction / routine names provided by the user whenever they are specified.  
3. If a name is missing, invent a concise, CamelCase name that matches the intent.  
4. Use as much Reactions language as possible and use Xtend only when necessary  
5. Do **not** wrap it in Markdown fences, and do **not** add commentary, explanations, or blank lines beyond what the language requires.
\end{lstlisting}
\end{promptbox}
\captionof{listing}{System Prompt for Code Generation (Reactions)}\label{lst:system_prompt_code_rl}
Based on the selected strategy options, the content of the corresponding files is loaded from a specified directory and incorporated into the final user prompt.
This aggregated prompt is then submitted to a user-selected \ac{LLM}, which generates the corresponding code and writes the output to the designated location.
This process is repeated for each queued task prompt. As with prompt generation, the prompt is divided into a system prompt and a user prompt.

Listing~\ref{lst:system_prompt_code_rl} shows the system prompt used for code generation in the Reactions language as a representative example.
The instructions embedded in the system prompt enforce adherence to the target \ac{MTL} grammar and are designed to minimize hallucination.





\subsection{Adaptation to other Model Transformation Languages}
\label{sub: ad}

Adapting this workflow to another MTL requires that the configuration of input and output files be set up appropriately. 
This includes the specification of the location of the new MTL's references, additional files required for e.g., few-shot prompting, and the desired save destination for the workflow's output. 
Whether the step of loading the model files~(metamodel artifacts) needs to be updated depends on whether the generated code is to be applied to the same model files. 
These configurations can be performed directly within the respective workflow nodes. 
Regarding adaptation steps that go beyond reconfiguring the locations of the input and output paths, the system prompt of the workflow must be adjusted to reflect the characteristics and capabilities of the desired \ac{MTL}. 
For instance, when applying the workflow to \ac{ATL}, \ac{ETL}, and \ac{QVTo}, the system prompts from the Reactions language, see Listings \autoref{lst:system_prompt_gen_rl} and \ref{lst:system_prompt_code_rl}, had to be updated. 
Furthermore, the improvement strategies, which are applicable for an \ac{MTL}, can vary and not all may be applicable. 
The Reactions language, for instance, included an additional improvement strategy for Xtext-based \ac{MTL}: including helper methods into prompt. Depending on the MTL's characteristics, further strategies could also be included.
To ensure that the semantics of the generated MTL code are validated appropriately, tests should be included. 
When applying the workflow to a new MTL, the manual effort required depends on whether pre-existing tests are used, must be written manually, or can be generated automatically. 
This also includes setting up the respective testing frameworks, which were different for the languages we investigated.

\section{Results and Evaluation}
\label{sec:evaluationandresults}



We evaluate which prompting strategies improve \ac{LLM}-generated \ac{MTL} code, considering the raw output of the \ac{LLM} using only a task prompt and model files (metamodel artifacts) without any additional context as the baseline.
The goal is to systematically evaluate \ac{LLM}-based code generation for \acp{MTL} across two dimensions: syntactic quality and semantic correctness.
We frame the evaluation around three evaluation questions (EQs) which operationalize the overarching Research Question posed in \autoref{sec:introduction}:

\begin{enumerate}[noitemsep, topsep=2pt]
  \item[\textbf{EQ1}] 
  To what extent do the prompting strategies improve the \emph{syntactic} quality of \ac{LLM}-generated \ac{MTL} code in terms of syntactic similarity and correctness compared to the baseline?

  \item[\textbf{EQ2}] 
  To what extent do the prompting strategies improve the \emph{semantic} correctness of \ac{LLM}-generated \ac{MTL} code compared to the baseline?

  \item[\textbf{EQ3}]
  To what extent does the choice of \ac{LLM} model affect the quality of the generated model transformation code across different \acp{MTL}?

\end{enumerate}

\subsection{Evaluation Metrics}


\paragraph{Syntactic Similarity}

To assess how closely the generated transformation code resembles the reference implementation, we employ the \textbf{\ac{ChrF} metric}~\cite{popovic_chrf_2015}, which measures character-level $F$-score between generated and reference code.
Unlike token-level metrics such as BLEU~\cite{papineni_bleu_2001}, ROUGE~\cite{lin_rouge_2004}, or METEOR~\cite{banerjee_meteor_2005}, \ac{ChrF} operates at the character level, making it more appropriate for programming languages where single characters such as semicolons or brackets carry syntactic significance.
Although \ac{ChrF} is not a reliable indicator of functional correctness, it provides a meaningful signal of how structurally divergent two code snippets are.

\paragraph{Syntactic Correctness}
\acp{LLM} may hallucinate invalid syntax or unknown tokens, yielding code that cannot be parsed.
The \textbf{Unparsed Rate}~\cite{bassamzadeh_comparative_2024} quantifies the proportion of generated snippets that fail parsing due to syntactic errors; lower values indicate that a larger share of the output is at least syntactically valid.
To obtain a finer-grained view of syntactic quality, we additionally report the \textbf{average number of parser-reported problems per line of code} ($PPL_s$) for each generated snippet~$s$.
This metric complements the Unparsed Rate by distinguishing between snippets that fail to parse entirely and those that contain only minor, Localized errors.
We compute $PPL_s$ by dividing the total number of parser-reported problems for snippet~$s$ by its non-empty, non-comment \ac{LoC}.
The parser infrastructure differs across \acp{MTL} to leverage the most accurate tooling available:
for the Reactions language and \ac{QVTo}, we use ANTLR-based parsers constructed from the respective grammars, 
for \ac{ATL}, we use the Eclipse ATL parser 
and for \ac{ETL}, we use the Eclipse Epsilon parser.

One limitation of $PPL_s$ is that reported problem counts depend on the error recovery mechanisms of the parsers that we use.
In particular, the ANTLR error-recovery mechanism may suppress subsequent errors once an earlier error triggers a recovery action.
Accordingly, the values we provide for $PPL_s$ are lower bounds.


\paragraph{Semantic Correctness}


To evaluate semantic correctness, we execute each generated snippet against concrete model instances and inspect the resulting output models.
We measure semantic quality using the \textbf{Pass@1} metric~\cite{paul_benchmarks_2024}, defined as the proportion of \emph{parsable} snippets for which the first generated candidate passes all associated test cases.
This metric is widely used in program synthesis and code-generation benchmarks~\cite{paul_benchmarks_2024}.



For the Reactions language, test cases operate on a \ac{VSUM}: some changes are applied to its constituent models, and the reactions file to preserve consistency.
For \ac{ATL}, \ac{ETL}, and \ac{QVTo}, separate test suites transform concrete source model instances.
In any case, we assert that the resulting target models are equal to the reference 
All test suites were \emph{manually written} to ensure that the semantic validation correctly reflects the intended transformation behavior.



\subsection{Experiment Setup}
\label{sec:experiment_setup}

We evaluate our approach on three \acp{LLM}: \textbf{\citet{gpt5_model_nodate}}, \textbf{\citet{noauthor_gemini_nodate}}, and \textbf{\citet{noauthor_claude_nodate}}. All models are treated as black-box generators, and no model parameters, weights, or training data are modified.

For each \ac{MTL} in the evaluation suite, every combination of prompting strategy and \ac{LLM} is executed on all available transformation scripts. The prompting strategies evaluated are: 
(i) \textbf{Baseline}: the task prompt together with the associated model files (metamodel artifacts) only; 
(ii) \textbf{Few-shots Prompting}: the baseline augmented with three representative few-shot examples; 
(iii) \textbf{Grammar Prompting}: the baseline augmented with the \ac{EBNF} grammar of the target \ac{MTL}; 
(iv) \textbf{Few-shots~+~Grammar Prompting}: the combination of few-shot and grammar prompting; and
(v) \textbf{Few-shots~+~Grammar Prompting~+~Helper Methods Inclusion}---the combination further augmented the prompt with predefined helper methods in the prompt (applicable only to the Reactions language, where Xtext-based utility functions are required).

\begin{table}[htbp]
\centering
\caption{Evaluation Suite statistics across the evaluated MTLs.}
\label{tab:db_stats}
\setlength{\tabcolsep}{5pt}
\renewcommand{\arraystretch}{1.15}

\begin{threeparttable}        \resizebox{\columnwidth}{!}{   \begin{tabular}{l r r r r r r r}                             
  \toprule
  & & \multicolumn{3}{c}{\textbf{LoC}}
  & \multicolumn{2}{c}{\textbf{Complexity (avg)}}
  & \\
  \cmidrule(lr){3-5}\cmidrule(lr){6-7}
  \textbf{Language}
  & \textbf{\#}
  & \textbf{Min}
  & \textbf{Max}
  & \textbf{Avg}
  & \textbf{\#R/H}
  & \textbf{MC}
  & \textbf{\#MM/M}
  \\
  \midrule
  Reactions  & 12 & 17 & 98  & 31.0 & 1.0 / 1.7 &  3.4 &  6/0 \\
  ATL        & 15 & 15 & 277 & 96.3 & 5.5 / 2.2 & 14.2 & 27/15 \\
  ETL        & 10 &  6 &  97 & 29.3 & 3.3 / 0.5 &  4.7 &  7/7 \\
  QVTo       & 10 & 13 &  25 & 18.5 & 3.7 / 0.0 &  3.8 & 1/0
  \\
\bottomrule
\end{tabular}
} 
\begin{tablenotes}
\footnotesize
\item[1] \# = number of transformation scripts.
\item[2] LoC = non-empty, non-comment lines of code. Min./Max. = LoC range.
\item[3] \#R/H = Avg.\ number of transformation rules / helpers per script
\item[4] MC = Avg.\ McCabe complexity per script. 
\item[5] \#MM/M = number of associated metamodel/model artifacts.
\end{tablenotes}
\end{threeparttable}
\end{table}
\subsection{Evaluation Suite}
\label{sec:evaluation suite}

To ensure a transparent and reproducible empirical evaluation across multiple \acp{MTL}, we construct an evaluation suite comprising executable transformation artifacts from four representative languages: Reactions, \ac{ATL}, \ac{ETL}, and \ac{QVTo}.
For each language, the evaluation suite includes:
(i)~executable transformation scripts serving as reference implementations,
(ii)~the corresponding input and output metamodels and model instances,
(iii)~manually written test suites that verify semantic correctness by executing the transformation and comparing the output against expected results,
and~(iv)~parser-based syntactic validation that counts parser-reported problems per snippet.


We draw artifacts from publicly available and official repositories \cite{vitruv-casestudies, mondo-atlzoo-benchmark, epsilon, eclipse-qvto}, prioritizing diversity in transformation intent (e.g., creation, update, mapping, refactoring) and selecting only executable, end-to-end examples.
Natural diversity in script complexity is preserved: \ac{ATL} examples tend to involve larger scripts and a higher number of associated model artifacts, whereas \ac{QVTo} and \ac{ETL} examples are more compact.
Furthermore, the Reactions language is defined for a specific transformation tasks, which is consistency preservation across multiple models~\cite{klareEnablingConsistencyViewbased2021}.
In contrast, \ac{ATL}, \ac{ETL} and \ac{QVTo} are general-purpose transformation languages, and more commonly used.
Thus, these languages are likely to be reflected in \ac{LLM} training data, whereas Reactions are unlikely to occur in such training data.







\autoref{tab:db_stats} summarizes the evaluation suite scale. We report the number of transformation scripts and code-size statistics in \ac{LoC}, measured as non-empty, non-comment lines.
We further report three complexity indicators, averaged per script, to characterize the transformation complexity of each task inspired by~\citet{gotz2021dedicated}: the number of transformation rules, the number of helpers, and the McCabe cyclomatic complexity calculated over all rules and helpers. We additionally report the number of associated metamodels and model artifacts, which also influence execution complexity and evaluation cost.









\subsection{Results and Findings}

\noindent To assess the statistical significance of the observed improvements, we employ a two-level testing procedure. The ANOVA statistic is defined in \citet{howell1992statistical}.

\paragraph{Overall significance across LLMs}

To evaluate whether different \acp{LLM} exhibit significantly different performance within a given prompting strategy, we used non-parametric tests appropriate for the metric type. For continuous metrics, including ChrF similarity (see~\autoref{tab:chrf}) and $PPL_s$ (see~\autoref{tab:error_rate}), we applied the \textbf{Friedman test} across \acp{LLM}. For binary outcomes, including parsability (see~\autoref{tab:unparsed}) and the combined event \emph{parsable and tests passed}, we used \textbf{Cochran’s Q test} across \acp{LLM}.
If one of the \acp{LLM} within a selected strategy (e.g., Few-Shot) produced significantly different results ($p < 0.05$), the triple is gray-shaded.

\paragraph{Pairwise comparisons against baseline} Individual comparisons against the baseline are conducted per \ac{LLM}, pairing each transformation script across strategies.
For \emph{binary} outcomes (parsability and test passed \& unpassed), we apply \textbf{McNemar's exact test}.
For \emph{continuous} metrics (\ac{ChrF} scores and $PPL_s$), we apply the \textbf{Wilcoxon signed-rank test}.
Results significantly better than the baseline ($p < 0.05$) are marked with~$^{*}$.

For clarity, the abbreviations used in the tables are as follows:  
\textbf{FS} represents few-shot prompting; \textbf{GR} represents grammar prompting; \textbf{HM} represents inputting some defined helper methods as prompt.


\subsubsection{Syntactic Evaluation (EQ1)}
\label{sec:syntactic_evaluation}
To answer EQ1, we evaluate syntactic quality using three complementary metrics: \ac{ChrF} for character-level similarity, the Unparsed Rate for parsability, and $PPL_s$ for fine-grained error density.

\paragraph{\textbf{ChrF}}
\begin{table*}[t]
\centering
\small
\setlength{\tabcolsep}{4pt}
\renewcommand{\arraystretch}{1.15}

\begin{tabular}{
l
ccc
ccc
ccc
ccc
}
\toprule
& \multicolumn{3}{c}{\textbf{Reactions}}
& \multicolumn{3}{c}{\textbf{ATL}}
& \multicolumn{3}{c}{\textbf{ETL}}
& \multicolumn{3}{c}{\textbf{QVTo}} \\
\cmidrule(lr){2-4}
\cmidrule(lr){5-7}
\cmidrule(lr){8-10}
\cmidrule(lr){11-13}

\textbf{Strategy}
& GPT & Gemini & Claude 
& GPT & Gemini & Claude 
& GPT & Gemini & Claude 
& GPT & Gemini & Claude \\
\midrule

Baseline     & \sig{0.62} & \sig{0.60} & \sig{0.78} & 0.59 & 0.57 & 0.58 & 0.64 & 0.61 & 0.61 & \sig{0.69} & \sig{0.73} & \sig{0.79} \\
Few-shot (FS)     & 0.87$^{*}$  & 0.85$^{*}$  & 0.88$^{*}$ & \sig{0.64$^{*}$} & \sig{0.58} & \sig{0.59} & 0.74$^{*}$ & 0.71$^{*}$ & 0.75$^{*}$ & \sig{0.80$^{*}$} & \sig{0.80$^{*}$} & \sig{0.81} \\
Grammar (GR)      & 0.80$^{*}$  & 0.78$^{*}$  & 0.85$^{*}$  & 0.59 & 0.57 & 0.58 & 0.63 & 0.63 & 0.62 & \sig{0.64} & \sig{0.70} & \sig{0.80} \\
FS + GR      & 0.86$^{*}$  & 0.88$^{*}$  & 0.87$^{*}$ & \sig{0.63$^{*}$} & \sig{0.57} & \sig{0.59} & 0.75$^{*}$ & 0.71$^{*}$ & 0.74$^{*}$ & \sig{0.80$^{*}$} & \sig{0.80$^{*}$} & \sig{0.83} \\
FS + GR + HM & 0.86$^{*}$  & 0.87$^{*}$  & 0.87$^{*}$  & -- & -- & -- & -- & -- & -- & -- & -- & -- \\

\bottomrule
\end{tabular}


\caption{Mean ChrF scores (Results significantly better than the baseline ($p < 0.05$) are marked with~$^{*}$. Higher is better). Gray-shaded triples within one strategy show significantly different results of the three LLMs.}
\label{tab:chrf}
\end{table*}

\autoref{tab:chrf} reports the mean ChrF scores across all four \acp{MTL}.
For the Reactions language, baseline values of GPT and Gemini are comparatively low, scoring 0.62 and 0.60, whereas Claude already achieves 0.78.
One plausible explanation is that Claude's training data contains structurally similar code, although this cannot be verified directly.
Across the remaining three \acp{MTL}, baseline scores vary: \ac{ATL} and \ac{ETL} exhibit moderate baselines (from 0.57 to 0.59 and from 0.61 to 0.64), while \ac{QVTo} shows stronger initial values (from 0.69 to 0.79), suggesting differing degrees of prior model exposure to these languages.

Few-shot prompting consistently yields the largest \ac{ChrF} gains across all four languages, with particularly pronounced effects for \ac{ETL} and Reactions (significantly better than the baseline for three \acp{LLM}).)
Grammar prompting alone provides smaller but consistent improvements, likely because it enforces correct syntactic constructs without conveying typical language idioms or usage patterns.
The FS~+~GR combination stabilizes performance at high levels across all models for \ac{ETL}, \ac{QVTo}, and Reactions.
For the Reactions language, adding helper methods (FS~+~GR~+~HM) results in only marginal changes in ChrF.

The Friedman test confirms statistically significant differences among three \acp{LLM} for \ac{QVTo} ($p < 0.05$), whereas differences for \ac{ETL} do not reach significance.


\begin{findingbox}{Finding 1}
Few-shot prompting is the primary driver of syntactic similarity across all four \acp{MTL}.
Grammar prompting alone provides moderate but consistent gains, and its combination with few-shot prompting (FS~+~GR) consistently achieves the highest \ac{ChrF} scores.
\end{findingbox}

\paragraph{\textbf{Unparsed Rate}}

\begin{table*}[t]
\centering
\small
\setlength{\tabcolsep}{4pt}
\renewcommand{\arraystretch}{1.15}

\begin{tabular}{
l
ccc
ccc
ccc
ccc
}
\toprule
& \multicolumn{3}{c}{\textbf{Reactions}}
& \multicolumn{3}{c}{\textbf{ATL}}
& \multicolumn{3}{c}{\textbf{ETL}}
& \multicolumn{3}{c}{\textbf{QVTo}} \\
\cmidrule(lr){2-4}
\cmidrule(lr){5-7}
\cmidrule(lr){8-10}
\cmidrule(lr){11-13}

\textbf{Strategy}
& GPT & Gemini & Claude 
& GPT & Gemini & Claude 
& GPT & Gemini & Claude
& GPT & Gemini & Claude \\
\midrule

Baseline
& 1.00 & 1.00 & 1.00
& 0.27 & 0.27 & 0.20
& \sig{0.30} & \sig{1.00} & \sig{0.10}
& 1.00 & 0.90 & 1.00 \\

Few-shot (FS)
& \sig{0.58} & \sig{0.75} & \sig{0.33$^{*}$} 
& 0.20 & 0.27 & 0.20
& \sig{0.10} & \sig{0.60} & \sig{0.10}
& 0.20$^{*}$ & 0.10$^{*}$ & 0.00$^{*}$ \\

Grammar (GR)
& \sig{0.92} & \sig{0.17$^{*}$}  & \sig{0.33$^{*}$} 
& \sig{0.60} & \sig{0.27} & \sig{0.13}
& \sig{0.30} & \sig{0.90} & \sig{0.20}
& \sig{0.40$^{*}$} & \sig{0.60} & \sig{0.90} \\

FS + GR
& 0.25$^{*}$  & 0.42$^{*}$  & 0.17$^{*}$ 
& 0.13 & 0.13 & 0.13
& \sig{0.20} & \sig{0.80} & \sig{0.20}
& 0.00$^{*}$ & 0.00$^{*}$ & 0.20$^{*}$ \\

FS + GR + HM
& \sig{0.17$^{*}$}  & \sig{0.5$^{*}$}  & \sig{0.08$^{*}$} 
& -- & -- & --
& -- & -- & --
& -- & -- & -- \\

\bottomrule
\end{tabular}

\caption{Average unparsed rate (Results significantly better than the baseline ($p < 0.05$) are marked with~$^{*}$. Lower is better). Gray-shaded triples within one strategy show significantly different results of the three LLMs.}
\label{tab:unparsed}
\end{table*}


\autoref{tab:unparsed} reports the average unparsed rate of generated snippets.
\textbf{For the Reactions language, the baseline unparsed rate is 1.00 across all three models, meaning that no baseline output is syntactically valid.} For \ac{QVTo}, the baseline is uniformly poor (from 0.90 to 1.00).
These results underscore the difficulty that \acp{LLM} face with low- or no-resource \acp{MTL}.
In contrast, \ac{ATL}, as a widely used \ac{MTL} with more public resources, exhibits a substantially lower baseline (from 0.20 to 0.27), consistent with its greater prevalence in publicly available code.
\ac{ETL} shows heterogeneous baseline behavior: Gemini produces no parsable output (1.00), whereas Claude achieves a rate of only 0.10.

Few-shot prompting substantially reduces the unparsed rate, particularly for \ac{QVTo}, where it drops to 0.00--0.20 across all models.
Grammar prompting alone has a limited effect for the Reactions language for GPT, likely because many critical constructs, such as Xtext XExpressions, are not explicitly captured by the grammar.
The FS~+~GR combination achieves the strongest reductions overall, bringing \ac{QVTo} to 0.00 for both GPT and Gemini.
For the Reactions language, the addition of helper methods (FS~+~GR~+~HM) yields further improvements, reducing the unparsed rate to 0.08--0.50 depending on the model.

The Cochran’s Q test confirms statistically significant differences among three \acp{LLM} for \ac{ETL} ($p < 0.05$).

\begin{findingbox}{Finding 2}
For low-resource \acp{MTL} such as the Reactions language and \ac{QVTo}, few-shot prompting, grammar prompting, and helper method inclusion provide the most significant gains.
For more widely known languages such as \ac{ATL} and \ac{ETL}, few-shot prompting alone is generally good to reduce the unparsed rate, even for the more complex transformations present in \ac{ATL}.

\end{findingbox}

\paragraph{\textbf{Average Number of Parser-reported Problems per Line of Code ($PPL_s$)}}

\begin{table*}[t]
\centering
\small
\setlength{\tabcolsep}{4pt}
\renewcommand{\arraystretch}{1.15}

\begin{tabular}{
l
ccc
ccc
ccc
ccc
}
\toprule
& \multicolumn{3}{c}{\textbf{Reactions}}
& \multicolumn{3}{c}{\textbf{ATL}}
& \multicolumn{3}{c}{\textbf{ETL}}
& \multicolumn{3}{c}{\textbf{QVTo}} \\
\cmidrule(lr){2-4}
\cmidrule(lr){5-7}
\cmidrule(lr){8-10}
\cmidrule(lr){11-13}

\textbf{Strategy}
& GPT & Gemini & Claude
& GPT & Gemini & Claude
& GPT & Gemini & Claude
& GPT & Gemini & Claude\\
\midrule

Baseline
& 0.0396 & 0.0562 & 0.0427
& 0.0092 & 0.0351 & 0.0138 
& \sig{0.0667} & \sig{0.0644} & \sig{0.0085}
& 0.3057 & 0.1674 & 0.1119 \\

Few-shot (FS)
& 0.0652 & 0.0335 & 0.0289
& 0.0024 & 0.0050 & 0.0072 
& \sig{0.0008} & \sig{0.0354} & \sig{0.0085}
& 0.0159$^{*}$ & 0.0100$^{*}$ & 0.0000$^{*}$ \\

Grammar (GR)
& \sig{0.0769} & \sig{0.0083$^{*}$} & \sig{0.0360}
& \sig{0.2757$^{*}$} & \sig{0.0204} & \sig{0.0005}
& 0.0412 & 0.0401 & 0.0122
& 0.1242 & 0.0784 & 0.0598 \\

FS + GR
& 0.0342 & 0.0437 & 0.0159$^{*}$
& 0.0019 & 0.0007 & 0.0007
& 0.0209 & 0.0324 & 0.0105
& 0.0000$^{*}$ & 0.0000$^{*}$ & 0.0099$^{*}$ \\

FS + GR + HM
& \sig{0.0145$^{*}$} & \sig{0.0546} & \sig{0.0160$^{*}$}
& -- & -- & --
& -- & -- & --
& -- & -- & -- \\

\bottomrule
\end{tabular}

\caption{Average number of syntax errors per line of code (LoC) (Results significantly better than the baseline ($p < 0.05$) are marked with~$^{*}$. Lower is better). Gray-shaded triples within one strategy show significantly different results of the three LLMs.}
\label{tab:error_rate}
\end{table*}


\autoref{tab:error_rate} reports the average number of parser-reported problems per line of code. 
For Reactions, the baseline error density ranges between 0.0396 and 0.0562. Few-shot prompting reduces the average number of errors for Gemini and Claude. Grammar prompting alone shows mixed effects: while Gemini benefits substantially (0.0083), GPT error density increases (0.0769). The combined FS~+~GR strategy stabilizes performance across models and significantly reduces errors for Claude (0.0159). Finally, the addition of helper methods (FS~+~GR~+~HM) yields the lowest error density for GPT (0.0145) and remains consistently low for Claude. 
For \ac{ATL}, the baseline error density is already low (from 0.0092 to 0.0351), indicating that the models generate largely well-formed \ac{ATL} code even without optimization. Few-shot prompting and the FS~+~GR combination further reduce this to near-zero levels. A notable outlier is grammar prompting alone for GPT on \ac{ATL} (0.2757), which significantly \emph{increases} the error rate compared to the baseline; this suggests that providing the grammar without accompanying examples may mislead certain models into producing syntactically aberrant constructs.
For \ac{ETL}, the baseline varies across models (0.0085 for Claude to 0.0667 for GPT). Few-shot prompting substantially reduces GPT's error rate to 0.0008, while the effect on Gemini and Claude is less pronounced. 
\ac{QVTo} exhibits the highest baseline error density (from 0.1119 to 0.3057), consistent with the high unparsed rates observed in \autoref{tab:unparsed}. Both few-shot prompting and the FS~+~GR combination yield significant reductions, with FS~+~GR bringing GPT and Gemini to 0.0000 errors per line. 

The Friedman test confirms statistically significant differences among three \acp{LLM} for \ac{ETL} ($p < 0.05$) generation with few-shot prompting and for Reactions and \ac{ATL} with grammar prompting, and for Reactions with the combination of few-shot and grammar prompting with heper method inclusion, whereas the differences for \ac{QVTo} do not reach significance.


\begin{findingbox}{Finding 3}
The combination of few-shot, grammar prompting, and helper method inclusion (for Reactions only) consistently achieves the lowest syntax error density across \acp{MTL}. Grammar prompting alone can be counterproductive for certain LLM models and language combinations.
\end{findingbox}


\subsubsection{Semantic Evaluation (EQ2)}
\label{sec:semantic_evaluation}

To answer EQ2, we now turn to semantic correctness, evaluating whether the parsable snippets produce functionally correct transformations.

\paragraph{\textbf{Pass@1}}


\begin{table*}[t]
\centering
\small
\setlength{\tabcolsep}{4pt}
\renewcommand{\arraystretch}{1.15}

\begin{tabular}{
l
ccc
ccc
ccc
ccc
}
\toprule
& \multicolumn{3}{c}{\textbf{Reactions}}
& \multicolumn{3}{c}{\textbf{ATL}}
& \multicolumn{3}{c}{\textbf{ETL}}
& \multicolumn{3}{c}{\textbf{QVTo}} \\
\cmidrule(lr){2-4}
\cmidrule(lr){5-7}
\cmidrule(lr){8-10}
\cmidrule(lr){11-13}

\textbf{Strategy}
& GPT & Gemini & Claude
& GPT & Gemini & Claude
& GPT & Gemini & Claude
& GPT & Gemini & Claude \\
\midrule

Baseline
& 0.00 & 0.00 & 0.00
& 0.67 & 0.67 & 0.67
& 0.20 & 0.00 & 0.10
& 0.00 & 0.10 & 0.00 \\

Few-shot (FS)
& 0.42 & 0.25 & 0.50$^{*}$
& 0.67 & 0.67 & 0.67
& 0.60 & 0.30 & 0.60
& 0.60$^{*}$ & 0.60 & 0.70$^{*}$ \\

Grammar (GR)
& 0.00 & 0.00 & 0.00
& 0.67 & 0.67 & 0.67
& 0.10 & 0.00 & 0.00
& 0.00 & 0.20 & 0.10 \\

FS + GR
& 0.50$^{*}$ & 0.42 & 0.50$^{*}$
& 0.67 & 0.67 & 0.67
& 0.50 & 0.10 & 0.50
& 0.50 & 0.70$^{*}$ & 0.60$^{*}$ \\

FS + GR + HM
& 0.58$^{*}$ & 0.42 & 0.42
& -- & -- & --
& -- & -- & -- 
& -- & -- & -- \\

\bottomrule
\end{tabular}


\caption{Pass@1 rate on syntactically parsable outputs (Results significantly better than the baseline ($p < 0.05$) are marked with~$^{*}$. Higher is better). Gray-shaded triples within one strategy show significantly different results of the three LLMs.}
\label{tab:test_cases}
\end{table*}



\autoref{tab:test_cases} reports the Pass@1 rate, computed exclusively on syntactically parsable snippets. 
For Reactions, the baseline Pass@1 is 0.00 because the unparsed rate is 1. Few-shot prompting yields improvements, increasing Pass@1 to 0.25--0.50, with statistically significant improvements for Claude. Grammar prompting alone does not improve semantic correctness. The combined FS~+~GR strategy further improves performance for Gemini and GPT. The addition of helper methods (FS~+~GR~+~HM) provides the highest Pass@1 rate for GPT (0.58). However, this improvement is not consistent across all \ac{LLM} models, as Gemini and Claude do not show additional gains and even decrease.
The reason for the marginal contribution of helper method inclusion might be that the few-shot examples serve to show how the helper methods work without having to explicitly include them. However, the primary challenge lies in semantics and reasoning about the transformation rules rather than predefined helper methods.



For \ac{ATL}, the Pass@1 rate remains at 0.67 across all strategies and models.
This result stands in contrast to \ac{ATL}'s strong syntactic performance (cf.\ \autoref{tab:chrf} and \autoref{tab:error_rate}), suggesting that while \acp{LLM} are capable of producing syntactically well-formed \ac{ATL} code, they struggle with the semantic complexity of the underlying transformation logic.
As shown in \autoref{tab:db_stats}, \ac{ATL} scripts in the evaluation suite are considerably longer and more complex than those of other languages (average 96~LoC vs.\ 18--31~LoC, average 14.2~McCabe complexity vs.\ 3.4--3.8~McCabe complexity), involving more intricate rule mappings and inter-rule dependencies.
The \acp{LLM} appears to handle the surface-level syntax of these longer scripts adequately, but fails to capture the deeper semantic intent.
For \ac{ETL}, the baseline Pass@1 is low (from 0.00 to 0.20), and few-shot prompting yields the most substantial improvement, raising Pass@1 to 0.30--0.60 depending on the model.
The FS~+~GR combination achieves comparable results (from 0.10 to 0.50), with GPT and Claude benefiting most.
\ac{QVTo} exhibits the most significant improvement: from a near-zero baseline (0.00--0.10) to 0.60--0.70 with few-shot prompting and 0.50--0.70 with FS~+~GR.
Grammar prompting alone yields negligible improvement, consistent with the pattern observed across all other metrics.

The Cochran’s Q test does not reach statistically significant differences among three \acp{LLM}.

\begin{findingbox}{Finding 4}
Few-shot prompting consistently provides the greatest improvements in Pass@1, while combining it with grammar prompting can further semantic correctness in some cases.
\textbf{Notably, syntactic proficiency does not imply semantic correctness}: for \ac{ATL}, the \acp{LLM} produce well-formed code but fail to capture the semantic complexity inherent in longer and more complex transformation scripts.
For \ac{QVTo}, few-shot prompting transforms the generated code from almost entirely non-functional to majority-correct.

\end{findingbox}

\subsubsection{Influence of LLM Choice (EQ3)}

This evaluation question explores whether the choice of \ac{LLM} influences the quality of generated model transformation code across \acp{MTL}.

The results show that the influence of \ac{LLM} choice is highly variable across \acp{MTL} and evaluation metrics. For \ac{QVTo}, the ChrF scores are significantly different between \acp{LLM} for all prompting strategies, reflecting constant differences in syntactic similarity (see gray shades in Table~\ref{tab:chrf}). Likewise, \ac{ETL} exhibits a constant and statistically significant difference between \acp{LLM} from parsability and $PPL_s$ across multiple strategies (see gray shades in Tables~\ref{tab:unparsed} and~\ref{tab:error_rate}), reflecting that syntactic correctness is sensitive to the LLM used for generation.
The Reactions Language shows statistically significant LLM differences mainly in parsability and $PPL_s$ for several strategies, reflecting a moderate sensitivity. In contrast, \ac{ATL} exhibits only isolated or no significant differences between LLMs across most strategies and metrics.

Despite these syntactic corrections and similarity differences, the Pass@1 rate does not show consistent statistically significant variation between \acp{LLM} across any \ac{MTL} (see gray shades in Table~\ref{tab:test_cases}), indicating that semantic correction is less dependent on the choice of \acp{LLM}.


\begin{findingbox}{Finding 5}
 LLM Model choice influences syntactic correctness and similarity, particularly for \ac{ETL} and \ac{QVTo} respectively, while its influence on semantic correction remains limited. 
 
\end{findingbox}

\section{Discussion}
\label{sec:discussion}
\subsection{Limitations}
This study is limited in language and model scope: four \acp{MTL} (Reactions language, \ac{ATL}, \ac{ETL} and \ac{QVTo}) are evaluated across three \acp{LLM} (GPT-5.1, Gemini 2.5-Pro, Claude Sonnet 4.5), which constrains generalizability, in particular for other low-resource languages. 
The strategy portfolio focuses on few-shot, grammar prompting, and helper method inclusion; more advanced techniques, such as \ac{RAG} or fine-tuning were not explored due to time and data constraints. The metric set is also constrained: ChrF is surface-based and pass@1 captures only the first attempt. Additionally, nondeterminism remains an inherent risk despite fixed parameters. 
The workflow is best characterized as structured and extensible rather than fully automated end-to-end, i.e., adaptation to a new MTL still requires language-specific manual effort, including system prompt redesign, resource curation, and test suite development. The current evaluation setup based on reverse-engineered prompts is systematic and reproducible, but may not fully represent real user prompt distributions, and whether the workflow performs equally well on underspecified or messy/misleading natural-language requests remains an open question.
A further limitation concerns potential training data contamination, as publicly available resources such as~\citet{mondo-atlzoo-benchmark} may have been included in the training corpora of the \acp{LLM}.

\subsection{Threats to Validity}
\paragraph{Internal Validity.} LLM non-determinism can shift results, and technical configuration (prompts, metamodel setup) influences all metrics. 
However, variations in prompt formulation or model temperature settings could still affect reproducibility across different experimental runs.

\paragraph{External Validity.} This study was evaluated on only four MTLs with limited dataset sizes, which restricts the generalizability of the results. The results may not represent the characteristics of all MTLs. Furthermore, we tested only three commercial LLMs, 
and the performance of open-source models or future models may differ. Future work needs to validate the effectiveness of the workflow on more MTLs and larger-scale datasets.

\paragraph{Construct Validity.}The metrics we adopted (ChrF, Unparsed Rate, Pass@1, etc.) may not fully capture all dimensions of code quality. For example, ChrF focuses on character-level similarity and may assign lower scores to functionally equivalent code with different structures. Pass@1 only evaluates the correctness of the first generation, without considering the potential for iterative refinement. Additionally, the coverage of test cases may affect the assessment of semantic correctness. We mitigate this by combining syntactic and semantic metrics for comprehensive evaluation, but a more complete quality assessment workflow remains to be explored. 

\paragraph{Conclusion Validity.} Statistical significance testing (ANOVA) confirmed differences among strategies, but the extremely poor baseline performance of the Reactions language (Unparsed Rate of 1.00) inflates the relative improvement of the strategies, though the absolute performance remains moderate (e.g., Pass@1 up to 0.33).
For ATL, improvements from certain strategies did not reach statistical significance or even showed no improvement (e.g., Pass@1), indicating that result stability varies across different MTLs. These findings suggest that the effectiveness of the workflow may depend on the characteristics of the target language. More broadly, the findings highlight a disconnect between syntactic and semantic improvement: Although prompt engineering can improve syntax correctness, especially for low-resource \acp{MTL}, it is inconsistent when it comes to the improvement of semantic correctness for more complex transformations. This is consistent with the bottleneck situations across different MTLs. For low-resource MTLs such as the Reactions language, the primary bottleneck lies in the LLM's insufficient syntax knowledge of the MTL, which can be largely mitigated through few-shot prompting; whereas for more widely used languages such as ATL, the bottleneck shifts to semantic and domain grounding, which current prompting strategies have yet to fully address.

\subsection{Workflow Generalizability and Extensibility}
The workflow's modular design facilitates adaptation to new MTLs. As 
demonstrated in \autoref{sub: ad}, extending to other \acp{MTL} primarily required adjusting 
system prompts and curating language-specific resources (grammar files, 
examples, test suites), while the core pipeline remained unchanged. This 
suggests that similar adaptations could apply to 
other MTLs, though the effectiveness may vary depending on language 
characteristics and available training data.
Beyond the current prompting strategies, the workflow can accommodate more 
advanced techniques. \ac{RAG} could augment 
prompts with dynamically retrieved domain knowledge, addressing the challenge 
of limited MTL training data. Fine-tuning, while resource-intensive, may 
become viable as more high-quality MTL datasets emerge. Additionally, 
automating parser creation and test execution would further reduce manual 
effort, making the workflow accessible to a broader range of MTL practitioners.
\section{Conclusion and Future Work}
\label{sec:conclusion}
This paper presented \textit{LLM4MTLs}, an automated and reproducible workflow for improving the reliability of \ac{LLM}-generated \ac{MTL} code and systematically evaluating its quality, together with an evaluation suite spanning four \acp{MTL} and an empirical evaluation on that suite.
Rather than modifying or fine-tuning \acp{LLM}, the workflow treats each LLM as a black-box code generator and focuses exclusively on prompt construction as the lever for improvement.
It provides an automated pipeline that standardizes prompt generation, code generation, and evaluation across \acp{MTL}. Using this workflow, we conducted an empirical evaluation on an evaluation suite spanning the Reactions language, \ac{ATL}, \ac{ETL}, and \ac{QVTo}, with different prompting strategy combinations across three \acp{LLM}, assessing both syntactic quality and semantic correctness. 
The few-shot prompting is the primary driver of syntactic improvement, though its ability to improve semantic correctness diminishes for complex transformations, such as \ac{ATL}.
Grammar prompting stabilizes generation when combined with few-shot examples, but can be counterproductive in isolation.
Finally, \ac{LLM} Model choice influences syntactic correctness and similarity, particularly for \ac{ETL} and \ac{QVTo} respectively, while its influence on semantic correctness remains limited across all \acp{MTL}.

\paragraph{Future Work}
Several directions emerge from this work. First, we plan to explore the use of \emph{agentic AI} approaches that incorporate iterative refinement loops, where the \ac{LLM} receives parser or test results feedback and refines its output autonomously. Second, \emph{\ac{RAG}} could dynamically select the most relevant examples and grammar fragments for a given task. Additionally, \emph{grammar-constrained decoding} offers a promising direction for eliminating parse errors. 
Another direction avenue to pursue includes qualitative analysis of semantic errors. While the present approach evaluates the generated transformation to see if the output models match the correct semantics, there is no investigation as to why these output models fail, for instance, whether model elements are missing or structurally misplaced. By conducting such an analysis, more actionable feedback for targeted prompt refinement could be gained.

\section*{Acknowledgements}
This paper is funded by the Deutsche Forschungsgemeinschaft (DFG, German Research Foundation) – SFB 1608 – 501798263. Thanks to our Textician Dan Shea.

\bibliography{software}
\section*{About the authors} 

\shortbio{Bowen Jiang}
{is a PhD researcher at Karlsruhe Institute of Technology (KIT), Germany. She received her M.Sc. from WASEDA University, Japan. Her research interests include MDE, software testing, and AI4SE. \authorcontact[https://mcse.kastel.kit.edu/staff_bowen_jiang.php]{bowen.jiang@kit.edu}}

\shortbio{Nathan Hagel}{is a PhD researcher at KIT, Germany. His research interests include MDE, uncertainty management for cyber-physical systems, DSLs, and cloud systems.
\authorcontact[https://mcse.kastel.kit.edu/staff_nathan_hagel.php]{nathan.hagel@kit.edu}}

\shortbio{Haowei Cheng}{is a PhD student at Waseda University, Japan.}
\authorcontact[https://haowei614.github.io/]{haowei.cheng@fuji.waseda.jp}

\shortbio{Benedikt Jutz}{is a PhD researcher at KIT. 
His research interests include concurrent editing techniques for multiple models and DSLs.
\authorcontact[https://dsis.kastel.kit.edu/staff_benedikt_jutz.php]{benedikt.jutz@kit.edu}}

\shortbio{Arne Lange}{is a Postdoctoral researcher at KIT. His research interests include multi-level modeling and model consistency preservation.
\authorcontact[https://dsis.kastel.kit.edu/staff_arne_lange.php]{arne.lange@kit.edu}}

\shortbio{Weixing Zhang}{is a PostDoc at KIT. He received his PhD at the University of Gothenburg. His research interests include SE, Empirical SE, AI4SE.\authorcontact[https://wilson008.github.io/]{weixing.zhang@kit.edu}}

\shortbio{Rahul Sharma}{is a PostDoc at Karlsruhe Institute of Technology. 
\authorcontact[https://dsis.kastel.kit.edu/staff_rahul_sharma.php]{rahul.sharam@kit.edu}}

\shortbio{Ralf Reussner}{is a professor at KIT since 2006. His research
interests include software architecture, predictable software quality, and view-based design methods for software-intensive
technical systems.
\authorcontact[https://dsis.kastel.kit.edu/staff_ralf_reussner.php]{reussner@kit.edu}}

\shortbio{Anne Koziolek}{{is a professor at KIT, Germany. She received her PhD degree from KIT in 2011. She is interested in MDE and agile development processes.
\authorcontact[https://mcse.kastel.kit.edu/staff_Koziolek_Anne.php]{koziolek@kit.edu}}}

\end{document}